\documentclass[10pt, prb, twocolumn]{revtex4}
\usepackage{epsfig}
\begin{document}

\title{Temperature and Field Dependence of the Mobility
in Liquid-Crystalline Conjugated Polymer Films \\ }
\author {S. J. Martin}
\affiliation{Department of Physics, University of Bath,
Claverton Down, Bath BA2 7AY, UK}
\author{A. Kambili}
\affiliation{Department of Physics, University of Bath,
Claverton Down, Bath BA2 7AY, UK} \affiliation{Institut
fuer Theoretische Physik, Universitaet Regensburg,
Universitaetsstrasse 31, Regensburg D-93053, Germany}
\thanks{Correspondence address \\
email address: agapi.kambili@physik.uni-regensburg.de}
\author {A. B. Walker}
\affiliation{Department of Physics, University of Bath,
Claverton Down, Bath BA2 7AY, UK}

\begin{abstract}
The transport properties of organic light-emitting
diodes in which the emissive layer is composed of
conjugated polymers in the liquid-crystalline phase
have been investigated. We have performed simulations
of the current transient response to an illumination
pulse via the Monte Carlo approach, and from the
transit times we have extracted the mobility of the
charge carriers as a function of both the electric
field and the temperature. The transport properties
of such films are different from their disordered
counterparts, with charge carrier mobilities
exhibiting only a weak dependence on both the
electric field and temperature. We show that for
spatially ordered polymer films, this weak
dependence arises for thermal energy being
comparable to the energetic disorder, due to the
combined effect of the electrostatic and thermal
energies. The inclusion of spatial disorder, on the
other hand, does not alter the qualitative behaviour
of the mobility, but results in decreasing its
absolute value. 
\end{abstract}
\maketitle
\pagestyle{empty}

\section{Introduction}
\par  Conjugated polymers, which derive their
semiconducting properties from delocalized $\pi$
bonding along the polymer chain \cite{pope}, enable
cheap, lightweight, portable and large area displays,
with promising applications to solar cells and
photodetectors \cite{gattinger, peumans, zhang}. Since
the first observation of polymer light-emitting diodes
(LEDs) \cite{burroughes}, detailed investigation of the
aspects of chemistry, physics and engineering of
materials has resulted in rapid progress \cite{hutten}.
In order to commercialize polymer devices, high
efficiencies, brightness, and carrier lifetimes are
required \cite{friend}. It is, therefore, essential to
fully understand the fundamental physics of electrical
transport through conjugated polymers, for a recent
review see \cite{walker}.
\par Previous theoretical studies have dealt with
strongly disordered organic materials, in which charge
transport is mainly attributed to hopping \cite{bassler,
novikov, rakhmanova}. The charge carrier mobilities of
such systems are low and exhibit a strong temperature
and electric field dependence. However, large carrier
mobilities are generally highly desirable. Most
recently, measurements in light-emitting diodes
\cite{redecker1, redecker2, martens} have demonstrated
enhanced carrier mobilities, varying only weakly with
the electric field, for devices characterized by high
degree of order in the polymeric material. This weak
dependence, which contradicts the theoretical
predictions, has been attributed to the purity of the
polymer films, but it is not yet understood.
\par This realization has motivated us to investigate
the transport properties of liquid-crystalline
conjugated polymer films in which the chains are
nematically aligned perpendicular to the direction of
transport. In a previous work \cite{kambili} we looked
into the character of transport through such polymer
films. By employing the Monte Carlo technique, we
showed that it is possible to obtain non-dispersive
transport in such systems. This observation is in
agreement with time-of-flight experiments conducted
on liquid-crystalline polyfluorene films
\cite{redecker1, redecker2}, which demonstrate
non-dispersive hole transport with enhanced charge
carrier mobilities compared to previously examined
conjugated polymers. Moreover, we established the
conditions under which such transport is retained.
However, our initial model was a simplistic one,
with the film morphology reduced to occupied sites
of a two-dimensional lattice, the variation of the
hopping probability with the field being rather
crude, and there was no temperature dependence. In
order to accomplish more detailed studies, we have
first extended the model to fully describe the
geometry of the polymer film, so that various film
morphologies, similar to those appearing in realistic
systems, can be considered. Secondly, we have
explicitly included all parameters of interest, such
as the electric field, temperature, and disorder (both
spatial and energetic).
\par The aim of the present work is to probe in detail
the field and temperature dependence of the mobility
in liquid-crystalline conjugated polymer films. In
particular, we discuss the effect of the electric
field on the inter-chain mobility of such polymer films,
and we attempt to explain the weak dependence on the
field. The interplay between electric field and
temperature on the transport characteristics is also
examined. We begin our investigation with spatially
ordered films in which all polymer chains are perfectly
aligned perpendicular to the direction of the field.
The additional effect of spatial disorder in the
film configurations on the transport properties of such
films, is also considered. To account for the chemical
regularity and the extended backbone conjugation of
liquid-crystalline chains we have included only a small
amount of energetic disorder.

\section{The Model}
\par In order to investigate the transport properties
of charge carriers moving within a liquid-crystalline
polymer film, we have performed numerical simulations
of the time-of-flight technique. The system under study
contains a polymer film sandwiched between two
electrodes, and charge carriers are generated on one
side by illumination of the electrode with an intense
pulse of light of short duration. The photogenerated
carriers move within the bulk of the film under the
effect of an external bias, whose sign determines the
type of the charge carriers that will get transported.
\par The film is composed of conjugated polymer chains
of length $L=100$nm, which are nematically aligned
perpendicular to the direction of the electric field,
chosen as the $x$ direction. For the construction of the
film periodic boundary conditions have been applied along
the other two directions. Based on the extended backbone
conjugation of liquid-crystalline polymers, such as
polyfluorene, that makes them stiff, and on bond
vibrations being of very high frequency and low
amplitude, we have assumed that the polymer chains can
be described as rigid rods. The thickness of the film
in the direction of transport has been taken equal to
$d=1\mu$m, since relatively thick films are required for
such measurements.
\par Hopping motion of the charge carriers under the
influence of the electric field is assumed, which in
general can be either intra- or inter-chain. This is
described by the Monte Carlo technique, in which the
carriers perform random walks within the bulk of the
film until they reach the collecting electrode. The
probability of hopping between two sites $i$ and $j$
is equal to

\begin{equation}
\mathcal{P}_{ij}= p_{ij}/\sum_{i\neq j}p_{ij}
\label{norm}
\end{equation}
where $p_{ij}$ is the unnormalized hopping probability,
which is taken to be of the form

\begin{equation}
p_{ij}=\gamma \exp{({-\frac{\varepsilon_{j}-\varepsilon_{i}-%
e\mathbf{E} \cdot\mathbf{r_{ij}}}{2K_{B}T}})}
\label{symm}
\end{equation}
$\mathbf{E}$ is the electric field, $\mathbf{r_{ij}}$
is the relative position vector, $K_{B}$ is the
Boltzmann factor, $T$ is the temperature, and $\gamma$
denotes the electronic wavefunction overlap. Here,
we have considered only nearest-neighbouring hopping
within a cutoff distance equal to 10{\AA}. $\varepsilon_{i}$ is the
on-site energy of site $i$ and is taken from a
Gaussian distribution

\begin{equation}
\rho(\varepsilon)=\frac{1}{\sqrt{2\pi\sigma^{2}}}\exp{(-\varepsilon^{2}/2\sigma^{2})}
\label{gauss}
\end{equation}
whose width $\sigma$ determines the degree of energetic
disorder present in the film. Notice that in
Equation \ref{symm} there is no other activation
energy except the difference in the electronic site
energies a carrier has to overcome during each jump.
Equation \ref{symm} satisfies the principle of
detailed balance \cite{yu}.
\par Intra-chain charge motion is, in general, a
very rapid process compared with the inter-chain one
\cite{stoneham}. Moreover, it was recently shown that
in liquid-crystalline polymer films inter-chain
hopping accounts for non-dispersive transport
\cite{kambili}. Taking these into account, we have
mainly focused on inter-chain transport, so that in
what follows, each Monte Carlo step corresponds to
the time required for an inter-chain jump.
\par The drift of the photogenerated carriers under
the external bias results in a time-dependent current
$I(t)$, which reads \cite{ferreira}

\begin{equation}
I(t)=\frac{\partial}{\partial t}\int_{0}^{d}{\rm d}x
\rho (x)(\frac{x}{d}-1)
\label{curr}
\end{equation}
$d$ is the film thickness, and $\rho(x)$ is the charge
density in the direction of transport, integrated over
the $y$ and $z$ directions. From the current transient
the transit time $t_{T}$, which is the time the carriers
need to reach the discharging electrode, is obtained.
We then use this extracted transit time to get the
mobility $\mu$ via the relation

\begin{equation}
\mu=\frac{d}{Et_{T}}
\label{mobil}
\end{equation}

\section{Results and Discussion}
\par
\begin{figure*}
\centerline{\psfig{figure=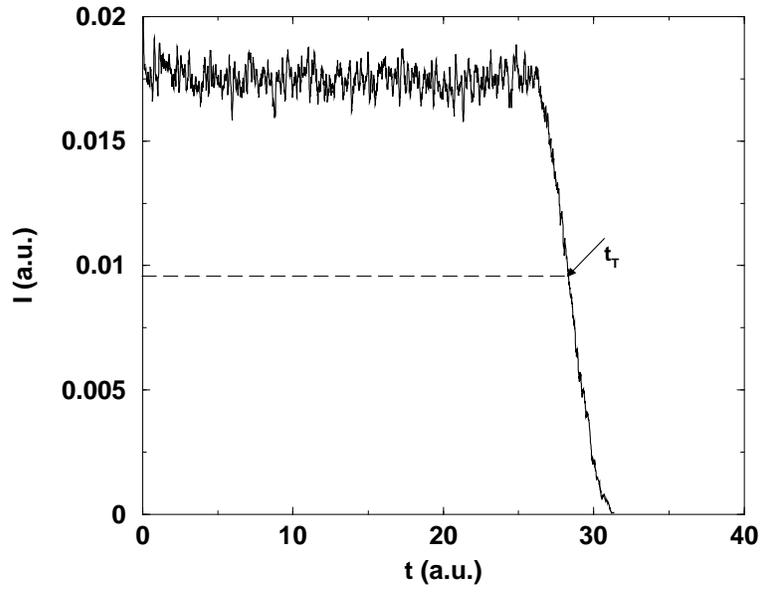,width=10cm}}
\caption{Current transient for energetic disorder $\alpha=2$,
with $E=3\times10^5$V/cm, and $T=300$K. The transit time $t_T$
is indicated by the arrow.}
\label{current}
\end{figure*}

 We present results for the charge carrier
mobilities of liquid-crystalline conjugated polymer
films, obtained by using the Monte Carlo model
described in the previous section. Unless stated
otherwise, all inter-chain distances are equal to 10{\AA}.
The output of such calculations is the current
transient, Eq. \ref{curr}, which arises from the
application of the external bias. In Fig.
\ref{current} we present a current transient from
our numerical simulations for electric field
$E=3\times10^5$V/cm at room temperature, and for disorder
$\alpha=2$, where $\alpha$ is defined as $\alpha=\sigma/K_BT_0$ and $T_0=300$K.
The calculated current exhibits the typical behaviour
of a time-of-flight signal, with a distinctive
plateau followed by a decaying tail, thus shows
non-dispersive transport \cite{scott}. From the
current plot we obtain the transit time $t_T$, which
is required for the calculation of the mobility, by
using the current integration mode. The latter gives
the transit time as the point at which the current
has fallen to half its value in the plateau region.
In Fig. \ref{current} the transit time is indicated
by the arrow.
\par
\begin{figure*}
\centerline{\psfig{figure=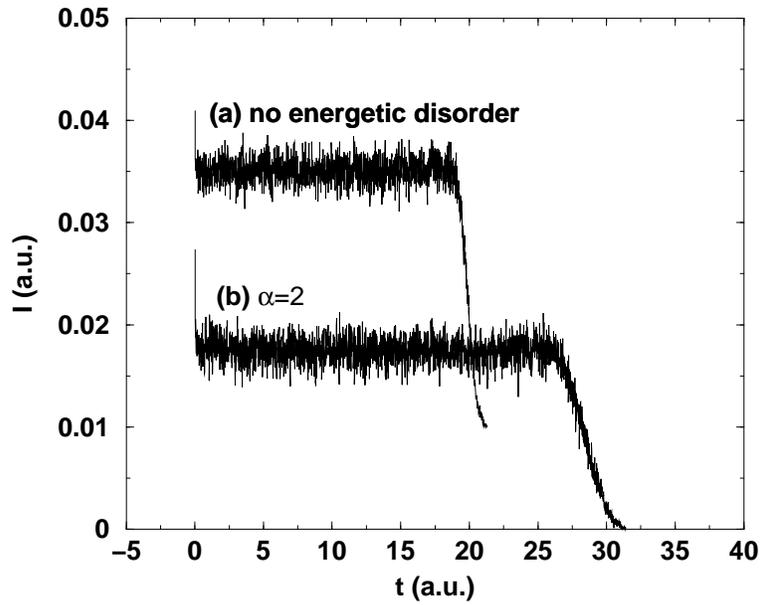,width=10cm}}
\caption{Current transient for $T=300$K and $E=3\times10^5$
V/cm. Curve (a) corresponds to the case of no energetic
disorder ($\delta\varepsilon=0$), and curve (b) corresponds to $\alpha=2$.
Curve (a) has been shifted along the $y$ axis for clarity.}
\label{peak}
\end{figure*}

 In most experimental time-of-flight signals,
the current transients initially exhibit a spike,
which rapidly falls into the plateau. However, in our
results such a pronounced peak is missing \cite{scher}.
We attribute the lack of the initial spike to the low
degree of disorder present in our system, and in
particular near the injecting electrode. A
demonstration of the effect of purity, in the initial
peak is presented in Fig. \ref{peak}. Curve (a) shows
the current transient for a film with no energetic
disorder at all ($\delta\varepsilon=0$), whereas curve (b) corresponds
to the case of $\alpha=2$. Curve (a) has been shifted along
the $y$ axis for clarity. After the illumination of
the electrode, the simulated photogenerated charge
packet begins its motion within the film. If no
disorder is present, the packet is adequately
described by a delta function moving within the bulk
of the film with constant mean velocity. If some
disorder is introduced into the system, the
propagating packet will become more extended
(Gaussian), and the carriers will no longer move
together. The resulting change in the charge profile
produces a spike in the current transient, as shown
in curve (b), even though the disorder that we have
considered is too small to give a pronounced peak.
Additionally, spatial irregularities near the
injecting electrode would lead to much larger peaks
\cite{kambili}, similar to those observed in
experiments.
\par
\begin{figure*}
\centerline{\psfig{figure=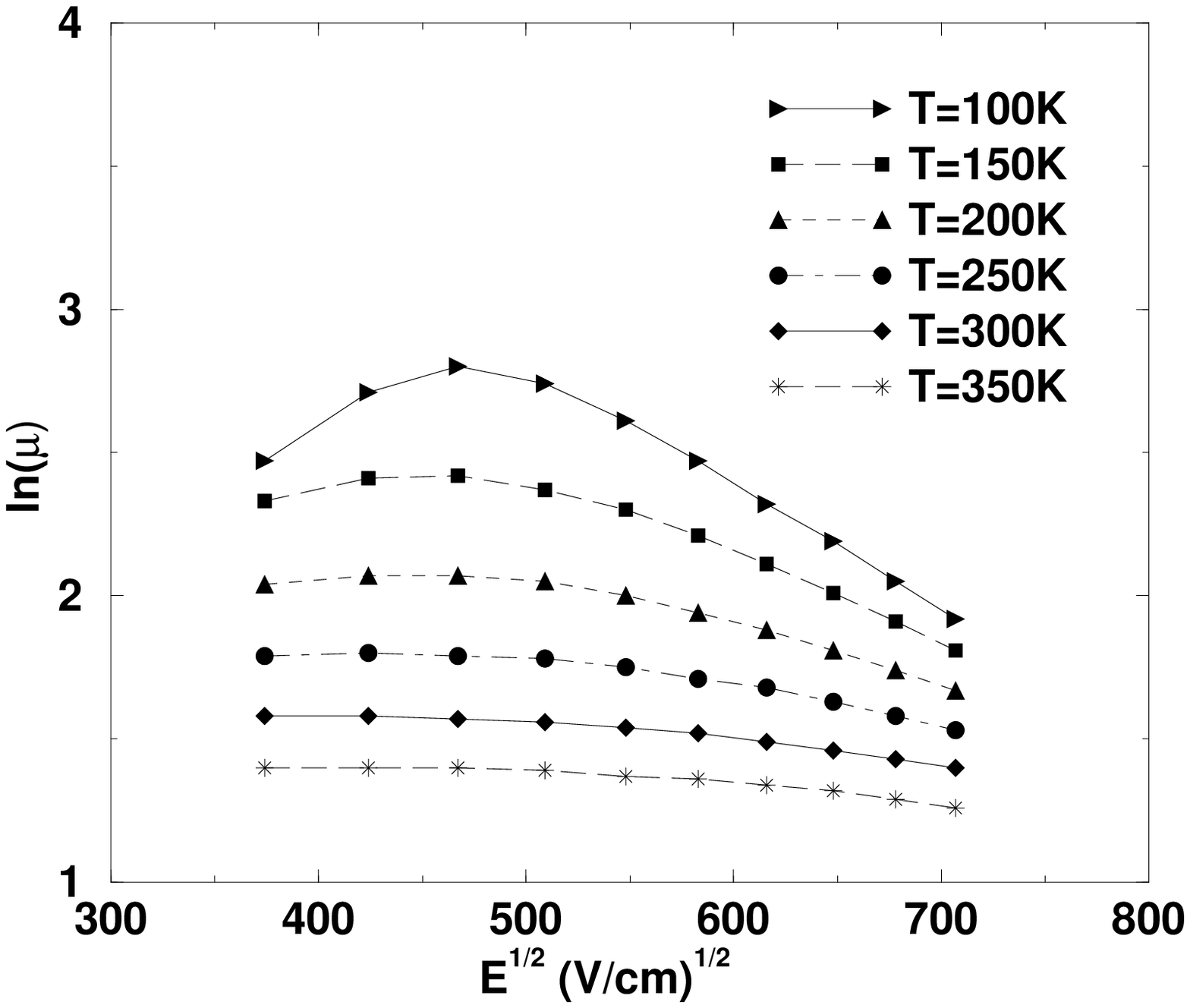,width=10cm}}
\caption{The logarithm of the mobility ln($\mu$) as a function
of $\sqrt{E}$ and for various temperatures $T$. The energetic
disorder is $\alpha=1$. $\mu$ is in arbitrary units.}
\label{mob1}
\end{figure*}

 By identifying the transit times from the
current transients for various values of the external
electric field and for different temperatures, we have
calculated the mobility by using Equation \ref{mobil}.
Figure \ref{mob1} shows one set of such results for
$\alpha=1$. In this graph the logarithm of the mobility,
ln($\mu$), has been plotted as a function of $\sqrt{E}$
for temperatures in the range of $100$K to $350$K. In
this Figure we see that for low temperatures $\ln{\mu}$
shows a more pronounced dependence on $\sqrt{E}$, with
a slight increase at low fields, which switches to a
decrease at larger values of the bias. As the
temperature increases, and, thus, the thermal energy
becomes comparable to the energetic disorder, this
variation disappears and the mobility changes only
weakly with the field. The reduction of the mobility
at large electric fields and at all temperatures is
the outcome of the saturation of the transit time,
since the electrostatic energy is sufficient for the
charge carriers to move fast enough and exit the film
at the minimum number of time steps required. However,
we should point out that we are not aware of any
experimental results which show such a decreasing
mobility. This indicates that in reality there are
not liquid-crystalline polymer films of such energetic
purity, even if they are characterized by perfect
alignment of the chains. 
\par
\begin{figure*}
\centerline{\psfig{figure=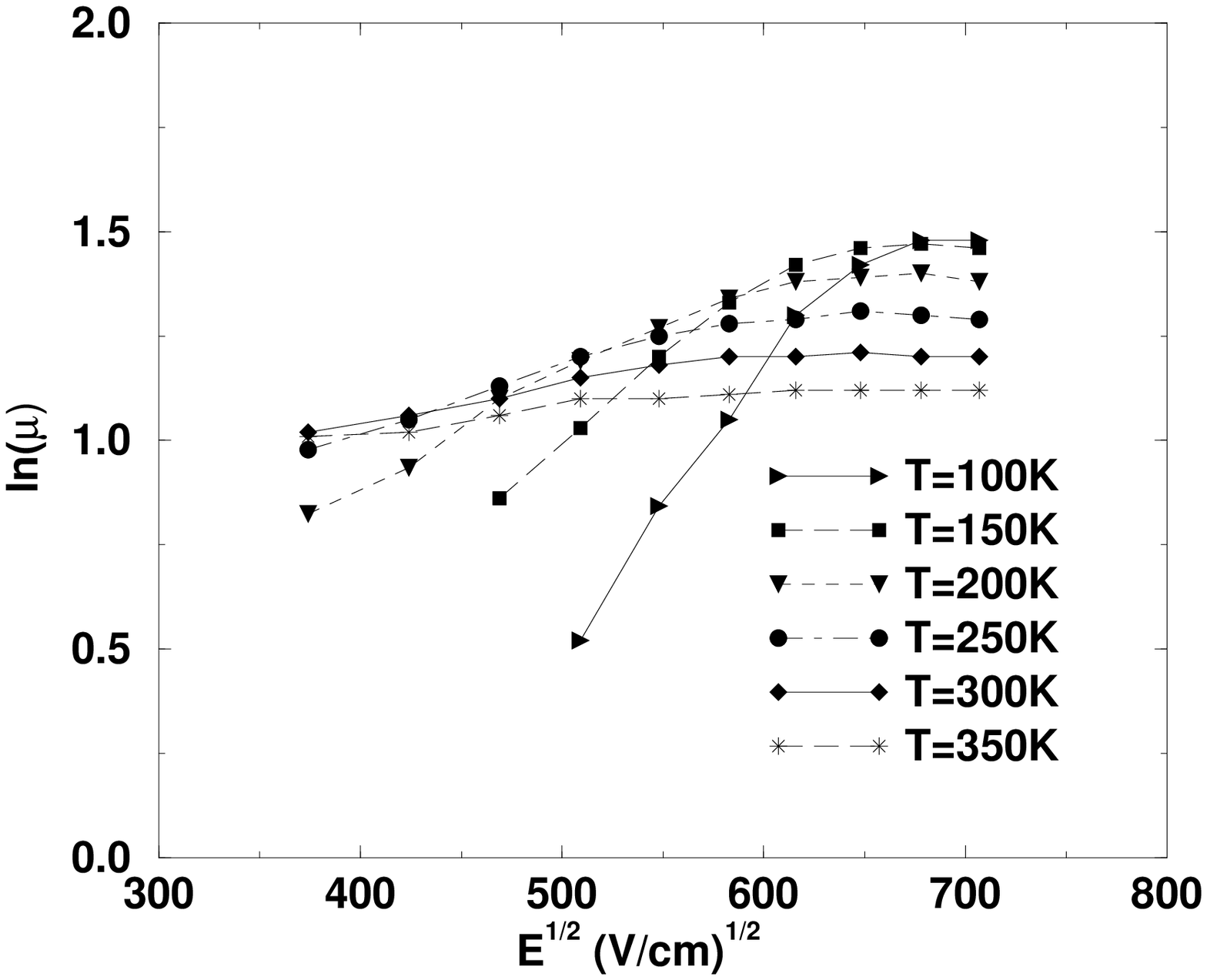,width=10cm}}
\caption{The logarithm of the mobility ln($\mu$) as a function
of $\sqrt{E}$ and for various temperatures $T$. The energetic
disorder is $\alpha=2$. $\mu$ is in arbitrary units.}
\label{mob2}
\end{figure*}

 In Figure \ref{mob2} the logarithm of the
mobility is shown as a function of $\sqrt{E}$ and for
the same range of temperatures as previously, but for
disorder $\alpha=2$. In contrast to Fig. \ref{mob1},
$\ln{\mu}$ now exhibits a positive gradient at all fields.
For small values of $T$ the thermal energy is not
adequate to enable the carriers to surmount the energy
barriers, and the electrostatic energy becomes
important. $\ln{\mu}$ increases with the field as the
electrostatic energy permits the carriers to follow
shorter paths within the film. Nevertheless, for larger
temperatures, the energetic disorder is rather small,
and the thermal and electrostatic energies have
similar contributions. Thus, the strong field
dependence seen in more disordered systems is smoothed
out and the carriers move with a constant velocity. A
direct comparison with experimental data
\cite{redecker1} shows that our calculations for the
mobility at room temperature and for $\alpha=2$ lead to
the same qualitative behaviour.
\par
\begin{figure*}
\centerline{\psfig{figure=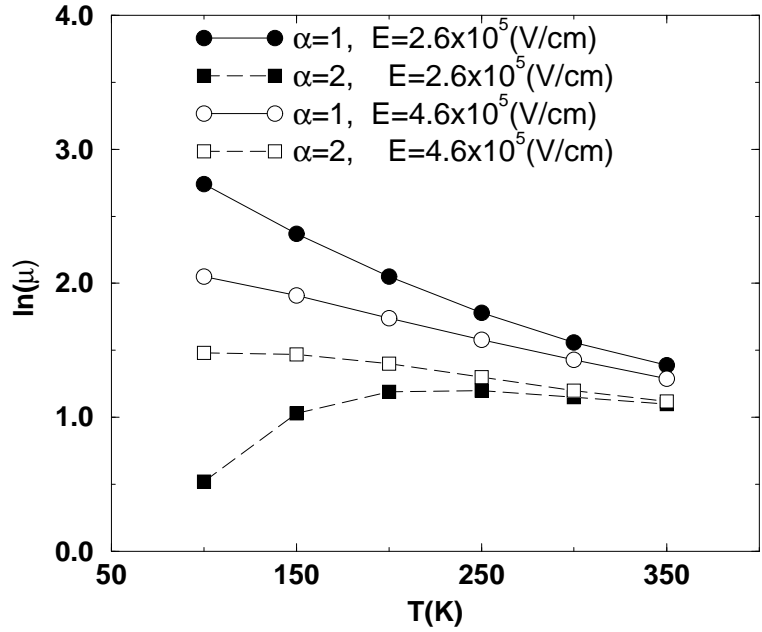,width=10cm}}
\caption{The logarithm of the mobility ln($\mu$) as a function of
$T$. The circles correspond to $\alpha=1$, and the squares to $\alpha=2$.
The filled symbols are for $E=2.6\times 10^5$V/cm, and the open symbols
are for $E=4.6\times 10^5$V/cm. $\mu$ is in arbitrary units.}
\label{temp}
\end{figure*}

 Figure \ref{temp} shows the behaviour of the
mobility as a function of temperature. The circles
correspond to $\alpha=1$ and the squares to $\alpha=2$. In the
former case, $\ln{\mu}$ decreases with $T$ for both
small (filled circles) and large (open circles)
electric fields. Energetic disorder is relatively
small so that the thermal energy dominates, making
the charge carriers to remain longer within the film,
following larger paths. Hence, the mobility decreases
with temperature. This is analogous to carriers being
scattered by phonons in conventional semiconductors.
When $\alpha$ increases, on the other hand, we see that at
a low electric field the mobility increases with
temperature (filled squares), whereas for a larger
field $\ln{\mu}$ decreases with $T$ (open squares). For
small fields, thermal activation is the leading
mechanism for the carriers to overcome the energy
barriers. Nevertheless, the energetic disorder is
still weak and at large fields longer paths due to
the excess thermal energy prevail, as before. This
is an important result because it shows how much the
mobility is affected by the interplay between
electrostatic and thermal energy in the case of small
energetic disorder. A similar behaviour of the
mobility as a function of temperature has also been
observed in other types of polymers within a Master
equation approach \cite{yu}, although in that case the
crossover appears to occur at the same field for all
temperatures considered. Moreover, at these small
amounts of energetic disorder changes in $\ln{\mu}$
with temperature are also weak.
\par
\begin{figure*}
\centerline{\psfig{figure=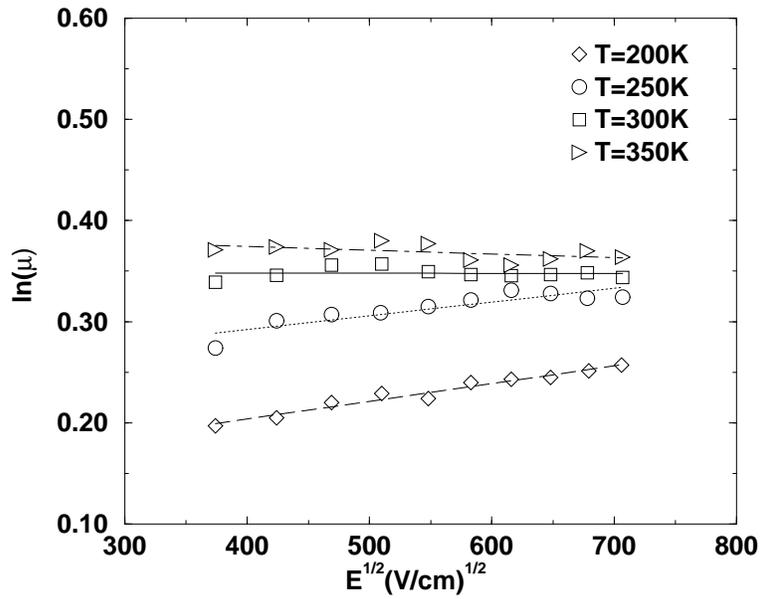,width=10cm}}
\caption{The logarithm of the mobility $\mu$ as a function of
$\sqrt{E}$ and for various temperatures $T$ for irregularly
spaced polymer chains. The lines are a guide to the eye. The
energetic disorder is equal to $\alpha=2$. $\mu$ is in arbitrary units.}
\label{irreg2}
\end{figure*}

 Nevertheless, this crossover from increasing to
decreasing mobility with temperature is not observed
if we depart from the equidistant chains arrangement.
In Fig. \ref{irreg2} we show $\ln\mu$ as a function of
$\sqrt{E}$ and for various temperatures, for a film
in which the liquid-crystalline chains are still all
nematically aligned, but at inter-chain distances that
vary randomly between 7{\AA} and 13{\AA} (irregularly spaced
chains). The disorder strength is $\alpha=2$, as in Figure
\ref{mob2}. A line has been fitted to the simulation
data points to guide the eye. Hopping is only allowed
between nearest-neighbours within the same cutoff
distance as before (10{\AA}), discussed in the previous
section. When the chains are irregularly spaced within
the film (Fig. \ref{irreg2}) the mobility is enhanced
with temperature for all values of the electric field,
as was shown in recent measurements of polyfluorene
films \cite{poplavskyy}. This behaviour resembles that
of strongly disordered polymer films \cite{hutten}.
However, in our case the increase of the mobility with
temperature arises from spatial irregularities in the
arrangement of the chains within the film. The
distribution in the relative position vector leads to
variations in the electrostatic energy (Eq. \ref{symm}),
so that thermal activation dominates, and the field
dependence of ln$\mu$ remains weak for all temperatures.
\par Comparison between Figures \ref{mob2} and
\ref{irreg2} also shows that, even though in both
regularly and irregularly spaced polymer films ln$\mu$
varies only slightly with the electric field, the
introduction of some spatial disorder leads to the
reduction of ln$\mu$. In the latter case, the charge
carriers require more time to reach the discharging
electrode as some nearest-neighbouring chains appear
in distances larger than 10{\AA}. Such jumps are
prohibited, and the carriers must find alternative
paths that might be longer. This would lead to greater
transit times and, therefore, to smaller mobilities,
as seen from Eq. \ref{mobil}.
\par
\begin{figure*}
\centerline{\psfig{figure=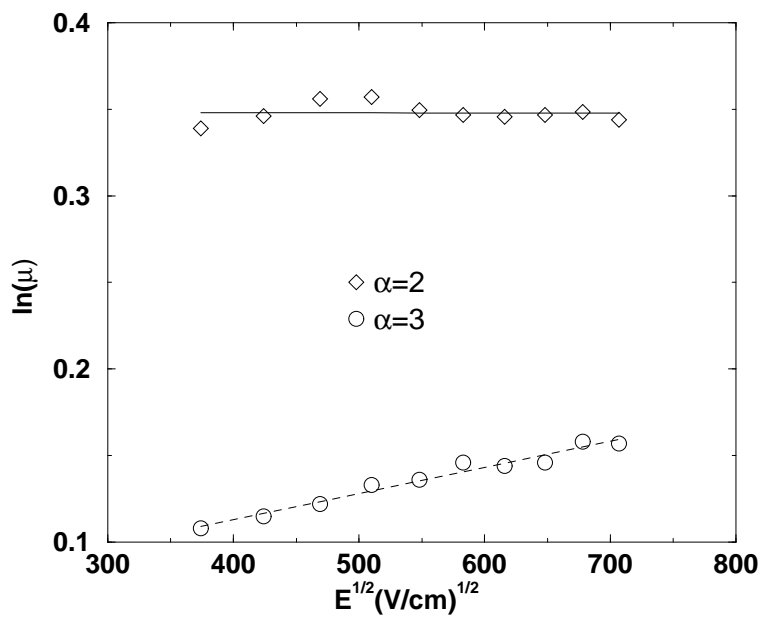,width=10cm}}
\caption{The logarithm of the mobility $\mu$ as a function of
$\sqrt{E}$ for irregularly spaced polymer chains, for $T=300$K.
The diamonds correspond to $\alpha=2$, and the circles to $\alpha=3$. The
lines are a guide to the eye. $\mu$ is in arbitrary units.}
\label{comp_irr}
\end{figure*}

 When the spatial disorder is retained and the
energetic disorder is increased, ln$\mu$ decreases even
more. Moreover, its qualitative behaviour approaches
the disordered limit, with a stronger dependence of
ln$\mu$ on the electric field. This is shown in Figure
\ref{comp_irr}, in which ln$\mu$ is plotted as a
function of $\sqrt{E}$ and at room temperature for an
irregularly spaced film, for two different strengths
of energetic disorder. The diamonds correspond to
$\alpha=2$, and the circles to $\alpha=3$. The lines are drawn
as a guide to the eye. Not only are the charge carriers
obliged to avoid large inter-chain distances, but
also they have to surmount larger energy barriers. In
this case, the charge carriers rely mostly on the
electrostatic energy as a source for the energy
required to overcome the barriers, similarly to
previously examined disordered films \cite{hutten}. 
\par From the above discussion it is understood that
the weak field and temperature dependence of ln$\mu$ in
liquid-crystalline conjugated polymer films originates
from the low energetic disorder present in the film.
The addition of spatial disorder, by irregularly
spacing the liquid-crystalline chains, does not affect
this dependence, but only results in the reduction of
the absolute value of the mobility. Nonetheless, it is
possible that another form of spatial disorder might
be present in such films, and this is the existence
of chains which deviate from perfect nematic alignment.
In other words, it is possible that during the formation
of such polymer films, some chains might appear at an
angle $\theta$ with respect to the axis of alignment.
We have examined this case  by assigning in our
simulations a random value of this angle $\theta$ to each
chain. The chains were also positioned in random
inter-chain distances, as in Figures \ref{irreg2} and
\ref{comp_irr}. 
\par
\begin{figure*}
\centerline{\psfig{figure=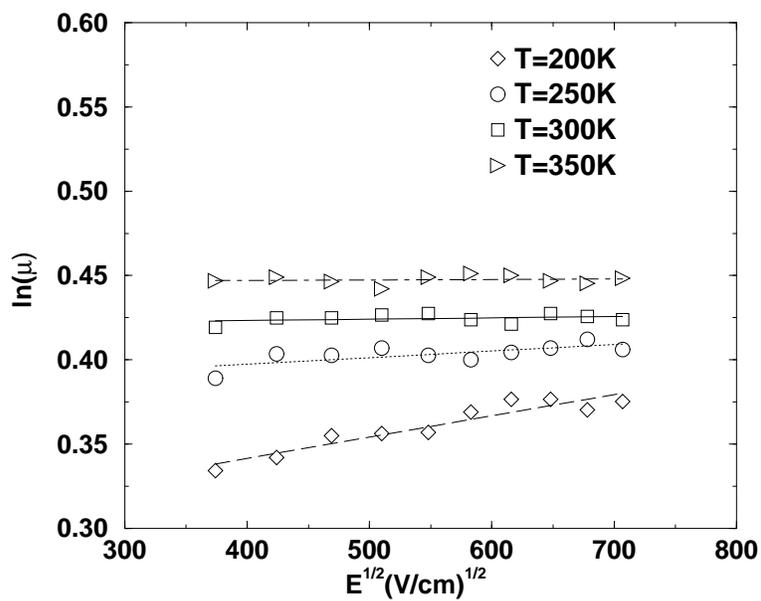,width=10cm}}
\caption{The logarithm of the mobility $\mu$ as a function of
$\sqrt{E}$ and for various temperatures $T$, for randomly
oriented and irregularly spaced polymer chains. The energetic
disorder is equal to $\alpha=2$. The lines are a guide to the eye.
$\mu$ is in arbitrary units.}
\label{mis5}
\end{figure*}

 Figure \ref{mis5} shows the outcome of such
calculations, in which $\theta$ varies randomly between $0^\circ$
and $5^\circ$. The logarithm of the mobility is plotted as a
function of $\sqrt{E}$, for various temperatures, as
before, with energetic disorder $\alpha=2$. The lines are
drawn as a guide to the eye. This Figure shows that
ln$\mu$ is only weakly dependent on the electric field,
and increases with temperature, in a similar fashion to
the case of perfect alignment. Hence, the inclusion of
an extra form of spatial disorder does not seem to
influence the qualitative behaviour of the mobility,
and stresses the importance of low energetic
disorder as the factor that determines the degree of
dependence on the field and the temperature.
Unfortunately, we cannot make any comparisons as to
whether the mobility increases or decreases with
respect to the perfect alignment case, since our
simulations for the randomly oriented case have been
performed for films with thickness $d=0.5\mu$m. Although
such film thickness might be considered small for real
time-of-flight experiments, we believe that our
calculations are able to represent the qualitative
behaviour of the mobility for films which show
deviations from perfect alignment.   

\section{Summary and Conclusions}
\par Liquid-crystalline polymer films are generally
expected to be characterized by spatial purity and
chemical regularity. In this paper, we have presented
calculations of the inter-chain mobility of charge
carriers through such polymer films, by employing the
Monte Carlo technique. In particular, we have
investigated the effect of low energetic disorder in
the electric field and temperature dependence of the
mobility, in order to account for the weak dependence
observed in experiments. Initially, films
with no spatial disorder were examined. Our numerical
simulations have shown that for relatively clean
systems, $\alpha=1$, the mobility decreases with the field,
as a result of the saturation of the transit time by
the electrostatic energy. Nevertheless, this behaviour
does not depict a realistic situation and a higher
amount of disorder, $\alpha=2$, was required to achieve
qualitative agreement with experiments
\cite{redecker1, redecker2, poplavskyy}. When the
thermal energy is not enough for the charge carriers
to overcome the energy barriers, the electrostatic
energy becomes important and ln$\mu$ increases with the
field. The weak field dependence of the mobility
arises from equivalent contributions from the thermal
and the electrostatic energy when the thermal energy
is comparable to the energetic disorder. 
\par The interplay between energetic disorder and
thermal energy also appears in the temperature
dependence of the mobility at a given electric field.
For low energetic disorder, $\alpha=1$, the mobility
decreases with temperature at all fields, as is
commonly seen in crystalline conductors. For larger
energetic disorder, $\alpha=2$, and for constant
inter-chain distances, the mobility increases with
temperature for small fields, but a crossover to the
opposite behaviour appears at larger fields. This
result is another demonstration of the combined effect
of the electric field and the temperature on the
mobility of charge carriers through spatially ordered
liquid-crystalline polymer films with low energetic
disorder.
\par When spatial disorder is taken into account, in
the form of randomly varying inter-chain distances,
the observed crossover from increasing to decreasing
mobility with temperature is no longer present.
Instead, the mobility always increases with $T$, as
was seen in the experiments \cite{poplavskyy}. Because 
of the presence of a distribution of the relative
position vectors, the effect of temperature is now
enhanced, even though the mobility is still only weakly
dependent on the electric field . Similar behaviour was
also observed in films in which the chains deviate
slightly from perfect nematic alignment perpendicular
to the direction of the electric field.
\par Finally, our numerical simulations in
liquid-crystalline polymer films have helped us to
distinguish between the effects of energetic and
spatial disorder in such systems. Low energetic
disorder, thus, high chemical purity of the chains, is
responsible for the weak dependence of the mobility in
the field. The inclusion of spatial disorder, associated
with irregularities in the arrangement of the polymer
chains within the film, does not make this dependence
any stronger so long as the energetic disorder remains
low. The influence of spatial disorder appears at the
quantitative  behaviour of the mobility, as it results
in decreasing considerably the absolute value of ln$\mu$
for a given range of fields and at a given temperature.
We should note, however, that our model does not allow
the calculation of the exact value of the mobility, or
make predictions about how many orders of magnitude the
mobility might increase or decrease. 
\par We thank Prof A. M. Stoneham for his useful comments
and discussions. The authors acknowledge the
{\it Engineering and Physical Sciences Research Council}
(EPSRC), and {\it Sharp Labs of Europe} for funding the
project. One of the authors, A. K., would also like to
thank the {\it Alexander von Humboldt} foundation.

\end{document}